\documentstyle[12pt,epsf,a4]{article}
\newcommand{\gsim}{\;\rlap{\lower 3.5 pt \hbox{$\mathchar \sim$}} \raise 1pt
 \hbox {$>$}\;}
\newcommand{\lsim}{\;\rlap{\lower 3.5 pt \hbox{$\mathchar \sim$}} \raise 1pt
 \hbox {$<$}\;}

\renewcommand{\thefootnote}{\fnsymbol{footnote}}
\newcommand{\logtwos}{L_{tW}}
\newcommand{\logtwms}{l_{tW}}
\newcommand{\logctheta}{l_\Theta}
\newcommand{\logmuW}{l_{\mu W}}
\newcommand{\yp}{y}

\begin{document}    

\title{\vskip-3cm{\baselineskip14pt
\centerline{\normalsize\hfill MPI/PhT/97--81}
\centerline{\normalsize\hfill TTP97--52}
\centerline{\normalsize\hfill hep-ph/9712228}
\centerline{\normalsize\hfill December 1997}
}
\vskip1.5cm
Complete Corrections of ${\cal O}(\alpha\alpha_s)$ to the Decay
of the $Z$ Boson into Bottom Quarks
}
\author{
 R.~Harlander$^{a}$,
 T.~Seidensticker$^{a}$
 and 
 M.~Steinhauser$^{b}$
}
\date{}
\maketitle

\begin{center}
$^a${\it Institut f\"ur Theoretische Teilchenphysik,
    Universit\"at Karlsruhe,\\ D-76128 Karlsruhe, Germany\\ }
  \vspace{3mm}
$^b${\it Max-Planck-Institut f\"ur Physik,
    Werner-Heisenberg-Institut,\\ D-80805 Munich, Germany\\ }
\end{center}
\vspace{1cm}

\begin{abstract}
  \noindent 
  For the vertex corrections to the partial decay rate
  $\Gamma(Z\to b\bar{b})$ involving the top quark only the leading terms
  of order $\alpha\alpha_s$ in the $1/M_t$ expansion are known.  In this
  work we compute the missing next-to-leading corrections.
  Thus at ${\cal O}(\alpha\alpha_s)$ the complete corrections to the
  decay of the $Z$ boson into bottom quarks are at hand.  \medskip

\noindent
PACS numbers: 12.38.Bx, 13.30.Eg, 13.38.Dg, 14.65.Fy
\end{abstract}

\thispagestyle{empty}
\newpage
\setcounter{page}{1}


\renewcommand{\thefootnote}{\arabic{footnote}}
\setcounter{footnote}{0}

\section{Introduction and notation}
\label{secint}

At the Large Electron Positron collider (LEP) at CERN approximately four
million decays of the $Z$ boson per experiment have been observed.
Because of this enormous statistic a lot of observables have been
measured with very high precision --- sometimes in the region of a few
per mille or even below.  Also the hadronic decay of the $Z$ boson has
been measured to an accuracy of roughly 0.1\% and amounts to
$\Gamma_{\rm had}=1742.4\pm 3.1$~MeV \cite{Alt97}.

From the theoretical side also a lot of effort has been undertaken in
order to give a precise prediction for $\Gamma_{\rm had}$. The QCD
corrections are known up to ${\cal O}(\alpha_s^3)$ both in the massless
limit \cite{GorKatLar91SurSam91} and at ${\cal O}(M_q^2/s)$
\cite{CheKue90,CheKue97}. A complete ${\cal O}(\alpha_s^2)$ result
\cite{KniKue89} and the leading terms at ${\cal O}(\alpha_s^3)$
\cite{CT94LRV9495} are available for the singlet contribution (for
reviews see also \cite{YelRep95,CKKRep}). 
Concerning purely electroweak
corrections a complete calculation is available at one-loop level
\cite{AkhBarRie86}, whereas at two loops the leading terms in $M_t$ are
known \cite{FleTarJeg93}.  The mixed corrections of ${\cal
  O}(\alpha\alpha_s)$ for the decay of the $Z$ boson into the quark
flavours $u, d, s$ and $c$ were considered in \cite{Kat92,CzaKue96,CzaMel97}.
From these results also those ${\cal O}(\alpha\alpha_s)$ contributions
to $\Gamma(Z\to b\bar{b})$ can be extracted where a photon or a $Z$ boson is
exchanged between the bottom quarks.  For the other class of diagrams
contributing to the decay into bottom quarks, namely those involving $W$
bosons and top quarks in the loop, only the leading $M_t^2$ corrections
\cite{FleJegRacTar92,CheKwiSte93} and the $\ln M_t^2$ terms
\cite{KwiSte95,Per95} are at hand.  In this work we close the gap and
provide the constant term at next-to-leading order.  In addition three
more terms in the high-$M_t$ expansion are computed.  It will be
demonstrated that these five terms provide a reliable result with
negligible errors.

It is useful to distinguish in the decay of the $Z$ boson
between universal and non-universal corrections.
Universal terms are independent of the produced fermions
and arise from corrections to the gauge boson propagators.
The non-universal corrections, sometimes just referred to as 
vertex corrections, depend on the fermion species considered.  
Of special interest is thereby the decay into bottom quarks
since in this case an additional dependence on the top quark mass
appears.

\begin{figure}[t]
\leavevmode
\centering
\epsffile[105 558 503 717]{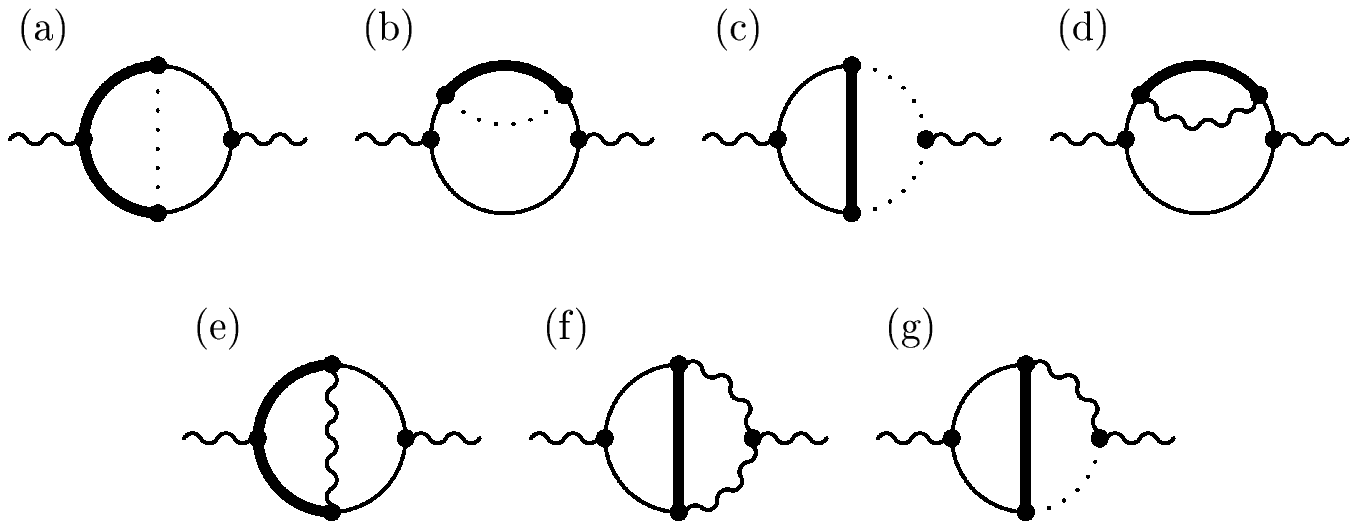}
\label{fig2l}
\caption{Diagrams contributing to $\delta\Gamma^W_b$ 
  in ${\cal O}(\alpha)$. Thin lines correspond to bottom quarks, thick
  lines to top quarks, dotted lines to Goldstone bosons and inner wavy
  lines represent $W$ bosons.}
\end{figure}

The partial decay rate of the $Z$ boson into bottom quarks can be
written in the form
\begin{eqnarray}
\Gamma(Z\to b\bar{b}) &=&
\Gamma^0
\left(
v_b^2 + a_b^2 \right)
+ \delta\Gamma_{\rm univ}
+ \delta\Gamma_b\,,
\label{eqzbb}
\end{eqnarray}
with $\Gamma^0=N_c M_Z \alpha/12 s_\Theta^2 c_\Theta^2$,
$v_b=-1/2+2s_\Theta^2/3$ and $a_b=-1/2$. $s_\Theta$ is the sine of the
weak mixing angle and $c_\Theta^2 = 1-s_\Theta^2$.
$\delta\Gamma_{\rm univ}$
represents the universal 
corrections and contains contributions from the transversal part
of the renormalized $Z$ boson polarization function evaluated for 
$q^2=M_Z^2$, $\Delta r$, the radiative corrections entering
the relation between $G_F, \alpha, M_Z$ and $M_W$, 
and the universal corrections arising from the $\gamma-Z$ interference.
$\delta\Gamma_b$ comprises all terms directly connected to the 
$Zb\bar{b}$ vertex. This includes both pure QED, QCD and electroweak 
corrections and the mixed contributions of order $\alpha\alpha_s$.
In this work we will concentrate on non-universal corrections
to $\delta\Gamma_b$ where the top quark is involved,
in the following denoted by\footnote{
  Although the index $W$ has been chosen, in a covariant gauge also the 
  diagrams involving the charged Goldstone bosons $\phi^\pm$ 
  have to be considered.}
$\delta\Gamma_b^W$.
The corresponding diagrams are pictured in Fig.~\ref{fig2l}.
To get the ${\cal O}(\alpha\alpha_s)$ corrections an internal gluon has
to be attached to them in all possible ways.
Analogously, we define $\delta\Gamma^Z_b$ as the contribution arising from 
the diagrams with internal $Z$ boson exchange. The mass of the
$b$ quark will be neglected throughout this paper.

Whereas $\delta\Gamma^Z_b$ is finite after the $Z$ boson contribution of
the wave function renormalization of the quarks is taken into account,
for $\delta\Gamma^W_b$ an additional counterterm induced by the Born and
${\cal O}(\alpha_s)$ result has to be added.  
In Feynman gauge it reads\footnote{ We are
  using dimensional regularization with space-time dimension
  $D=4-2\varepsilon$, suppressing, however, the $\ln 4\pi$ and $\gamma_E$
  terms in all formulas.} \cite{CzaKue96}:
\begin{eqnarray}
\delta\Gamma^W_b &=& \delta\Gamma^{0,W}_b + \delta\Gamma^{{\rm ct},W}_b,
\nonumber\\
\delta\Gamma^{ct,W}_b &=& \Gamma^0
  {1\over s_\Theta^2}{\alpha\over \pi} c_\Theta^2
  \bigg\{
           {1\over \varepsilon}\,\bigg[
               - {1\over 6} 
               - {1\over 3}\,c_\Theta^2
               + {\alpha_s\over \pi}\,C_F\,\left(
                   - {1\over 8} 
                   - {1\over 4}\,c_\Theta^2
                  \right)
              \bigg]
\nonumber\\&&\mbox{}
           - {5\over 18} 
           - {5\over 9}\,c_\Theta^2 
           + \left(
               - {1\over 3} 
               - {2\over 3}\,c_\Theta^2
              \right)\,\logmuW 
           + \left(
               - {1\over 6} 
               - {1\over 3}\,c_\Theta^2
              \right)\,\logctheta 
\nonumber\\&&\mbox{}
           + {\alpha_s\over \pi}\,C_F\,\bigg[
               - {55\over 48} 
               - {55\over 24}\,c_\Theta^2 
               + \left(
                     1 
                   + 2\,c_\Theta^2
                  \right)\,\zeta_3
\nonumber\\&&\mbox{}\hspace{1em}
               + \left(
                   - {3\over 8} 
                   - {3\over 4}\,c_\Theta^2
                  \right)\,\logmuW 
               + \left(
                   - {1\over 4} 
                   - {1\over 2}\,c_\Theta^2
                  \right)\,\logctheta 
              \bigg]
          \bigg\}\,,
\label{eqct}
\end{eqnarray}
with $\zeta_3 \approx 1.202056903$,
$l_{\mu W} = \ln(\mu^2/M_W^2)$ and $l_\Theta = \ln c_\Theta^2$.
$\delta\Gamma^{0,W}_b$ is the contribution from the Feynman diagrams
depicted in Fig.~\ref{fig2l}.

In principle there are several options to write down an equation for
$\Gamma(Z\to b\bar{b})$ containing the radiative corrections.
Eq.~(\ref{eqzbb}) is only one possibility.  Often parts of the
vertex corrections are comprised into so-called effective couplings and
the QCD corrections are described by a factor $(1+\alpha_s/\pi)$.
This will be discussed in Section~\ref{secres} where the results are
presented. In the next Section some details
on the calculation are given.

\section{The calculation}
\label{seccal}

In the following the method used for the calculation is briefly
described.  Instead of directly computing the two-loop vertex diagrams
the three-loop polarization function of the $Z$ boson is considered.
The imaginary part then immediately leads to contributions to
$\delta\Gamma_b^W$.  This avoids the separate treatment of infra-red and
collinear divergences and has the advantage that the advanced
computational tools available for two-point functions can be used.  In
addition, as the top quark is much heavier than all other mass scales
involved in the problem, it is tempting to perform an expansion in the
inverse top quark mass. The method which provides a systematic expansion
in $1/M_t$ is given by the so-called hard mass procedure (HMP)
\cite{hmp}.

The application of the HMP requires in a first step the identification
of all hard subgraphs which contain all lines carrying the large mass
and whose connectivity components are one-particle-irreducible with
respect to lines corresponding to light or massless particles.  These
subgraphs have to be expanded with respect to the small masses and the
external momenta.  In the full diagram all lines of the hard subgraph
have to be shrunk to a point to give the so-called co-subgraph. The
result of the expansion is inserted as an effective vertex. Finally the
loop integrations are performed.

At two-loop level seven diagrams have to be considered. Their
calculation would still
be feasible by hand. At three loops, however, 69 diagrams contribute
and a computation by hand is very painful
especially if higher order terms in the $1/M_t$ expansion are considered.
The HMP applied to the 69 initial diagrams contributing to $\delta\Gamma_b^W$
results in 234 sub-diagrams which have to be expanded in their
small quantities.
For this reason the program package EXP written in Fortran~90
was developed. It generates all possible subgraphs of a given diagram
together with some auxiliary files which allow the computation to 
be done automatically. The output of EXP can directly be used as input for
MATAD~\cite{Ste96} and MINCER~\cite{MINCER} where the loop integration
is performed.

It is possible to separate the diagrams at two-loop level 
(see Fig.~\ref{fig2l}) into two classes:
If the HMP is applied to the diagrams of type $(a)$, $(b)$, $(d)$ or $(e)$
the resulting integrals factorize into massive diagrams with only one 
mass scale times purely massless ones with external momentum $q$.
For the types $(c)$, $(f)$ and $(g)$ the HMP also leads to a 
factorization of the original integral. Here, however, massive one-loop
integrals with two different masses and non-vanishing external 
momentum have to be computed. At three-loop level, where the same
classification is valid, in principle
even the ${\cal O}(\varepsilon)$ part would be necessary.
For this reason we apply the HMP again to these
co-subgraphs with the conditions
$M_W^2\gg\xi_W M_W^2\gg q^2$, or alternatively, $\xi_W M_W^2\gg M_W^2\gg q^2$.
Here, $\xi_W$ is the gauge parameter appearing in the $W$ and $\phi$
propagators.
Of course, both descriptions must lead to identical final results as the
$\xi_W$ dependence drops out at the very end. In intermediate steps,
however, the expressions which have to be evaluated are different.

At the end $q^2=M_Z^2$ has to be chosen. Then the above inequalities are
seemingly inadequate. However, a closer look to the corresponding
diagrams shows that actually $(2M_W)^2$, respectively $(M_W + M_t)^2$, has
to be compared with $q^2$. The procedure is furthermore justified a
posteriori as rapidly converging series are obtained from the
expansions.


\section{Results}
\label{secres}

Since we are interested in the virtual effect of the top quark which 
renders the decay of the $Z$ boson into bottom quarks different from
the one into other down-type quarks, in the following the quantity
$\delta\Gamma_b^{0,W} - \delta\Gamma_d^{0,W}$ is considered. 
In this difference the counterterm contribution
exactly cancels. This means that the renormalized partial decay
rate can be written in the form
\begin{eqnarray}
\delta\Gamma_b^W&=&
\delta\Gamma_d^W
+ \left( \delta\Gamma_b^{0,W} - \delta\Gamma_d^{0,W} \right),
\end{eqnarray}
where $\delta\Gamma_d^W$ is the contribution 
from the diagrams involving a $W$ boson to the partial decay rate
$\Gamma(Z\to d\bar{d})$ \cite{CzaKue96}.
Note that besides the $1/\varepsilon$ poles also the dependence on $\xi_W$
drops out in the difference $\delta\Gamma_b^{0,W} - \delta\Gamma_d^{0,W}$.
For convenience the following notation is introduced:
\begin{eqnarray}
\delta \Gamma_{b-d}^{W} \,\,=\,\,
\delta \Gamma_b^{0,W} - \delta \Gamma_d^{0,W} &=& \delta \Gamma_{b-d}^{(1)}
 + {\alpha_s\over \pi} C_F \,\delta \Gamma_{b-d}^{(2)} \,.
\label{eqgamdiff}
\end{eqnarray}
For $\delta \Gamma_{b-d}^{(1)}$ we get:
\begin{eqnarray}
  \lefteqn{\delta \Gamma_{b-d}^{(1)} = \Gamma^0
    {1\over s_{\Theta}^2}{\alpha\over \pi} \times} 
\nonumber\\&&\mbox{}
\bigg\{
       {M_t^2\over M_W^2}\,\bigg[
           - {1\over 24} 
           - {1\over 48}\,{1\over \yp}
          \bigg]
\nonumber\\&&\mbox{}
       + \bigg[
         {857\over 3240} 
         + {407\over 2592}\,{1\over \yp}
         - {41\over 567}\,\yp 
         - {731\over 56700}\,\yp^2 
         - {11219\over 363825}\,\yp^3 
         + \logtwos\,\bigg(
             - {1\over 8} 
             - {1\over 18}\,{1\over \yp}
\nonumber\\&&\mbox{}
\hspace{1em}
             - {1\over 36}\,\yp 
            \bigg) 
         + \logctheta\,\bigg(
             {1\over 54}\,{1\over \yp}
             - {7\over 270}\,\yp 
             + {1\over 135}\,\yp^{2} 
             + {16\over 945}\,\yp^3 
            \bigg)
         + {\cal O} (\yp^4)
         \bigg]
\nonumber\\&&\mbox{}
       + {M_W^2\over M_t^2}\,\bigg[
             {749\over 12960} 
           + {61\over 432}\,{1\over \yp}
           - {10033\over 22680}\,\yp 
           + {947\over 22680}\,\yp^2 
           + {298\over 4455}\,\yp^3 
           + {11038\over 405405}\,\yp^4 
\nonumber\\&&\mbox{}
\hspace{1em}
           + \logtwos\,\bigg(
               - {23\over 72} 
               - {5\over 24}\,{1\over \yp}
               + {2\over 9}\,\yp 
               + {1\over 18}\,\yp^{2} 
              \bigg) 
           + {\cal O} (\yp^5)
          \bigg]
\nonumber\\&&\mbox{}
       + \left({M_W^2\over M_t^2}\right)^{2}\,\bigg[
           - {3707\over 6480} 
           + {43\over 288}\,{1\over \yp}
           - {15251\over 16200}\,\yp 
           + {92531\over 56700}\,\yp^2 
           + {137\over 22275}\,\yp^3 
           - {18464\over 135135}\,\yp^4 
\nonumber\\&&\mbox{}
\hspace{1em}
           - {20896\over 405405}\,\yp^5 
           + \logtwos\,\bigg(
               - {1\over 9} 
               - {17\over 48}\,{1\over \yp}
               + {19\over 20}\,\yp 
               - {5\over 9}\,\yp^2 
               - {2\over 15}\,\yp^3 
              \bigg) 
           + {\cal O} (\yp^6)
          \bigg]
\nonumber\\&&\mbox{}
       + \left({M_W^2\over M_t^2}\right)^{3}\,\bigg[
           - {22399\over 12960} 
           + {65\over 432}\,{1\over \yp}
           + {1063\over 7560}\,\yp 
           + {285163\over 48600}\,\yp^2 
           - {2425274\over 467775}\,\yp^3 
\nonumber\\&&\mbox{}
\hspace{1em}
           - {96794\over 868725}\,\yp^4 
           + {127808\over 405405}\,\yp^5 
           + {753776\over 6891885}\,\yp^6 
           + \logtwos\,\bigg(
                 {215\over 216} 
               - {109\over 216}\,{1\over \yp}
               + {43\over 27}\,\yp 
\nonumber\\&&\mbox{}
\hspace{1em}
               - {371\over 90}\,\yp^2 
               + {8\over 5}\,\yp^3 
               + {16\over 45}\,\yp^4 
              \bigg)
           + {\cal O} (\yp^7)
          \bigg]
\nonumber\\&&\mbox{}
       + {\cal O} \left( \left( {M_W^2\over M_t^2} \right)^4 \right)   
\bigg\}\,,
\label{eqzbb1loop}
\end{eqnarray}
with $\yp = 1/(4 c_\Theta^2)$, $\zeta_2 = \pi^2/6$, $L_{tW} =
\ln(M_t^2/M_W^2)$ and $l_{\Theta}$ as in Eq.~(\ref{eqct}).  Note that
the successive application of the HMP to the diagrams of class $(c),
(f)$ and $(g)$ (see Fig.~\ref{fig2l}) leads to an expansion in
$1/c_\Theta^2$. However, the coefficients decrease very rapidly which
justifies this strategy.  Moreover, a $W$ boson in the final state is always
accompanied by either another $W$ boson or a top quark.  For
this reason we choose the variable $\yp$ for the presentation of the
results. Nevertheless, two remarks concerning the expansion in
$1/c_\Theta^2$ are in order: First, note that the above series is
nested and therefore each order in $1/M_t^2$ produces an additional
power in $\yp$.  To demonstrate the quality of the convergence all
available terms are displayed.  Second, one recognizes that the
coefficients of the logarithms seem to be truncated series in
$1/c_\Theta^2$, which suggests that they might be exact.  These remarks
are also taken over to the ${\cal O}(\alpha\alpha_s)$ corrections given
by
\begin{eqnarray}
\lefteqn{ \delta \Gamma_{b-d}^{(2)} = \Gamma^0
    {1\over s_{\Theta}^2}{\alpha\over \pi} \times } 
\nonumber\\&&\mbox{} 
\bigg\{
           {M_t^2\over M_W^2}\,\bigg[
               - {1\over 32} 
               - {1\over 64}\,{1\over \yp} 
               + \zeta_2\,\bigg(
                     {1\over 16} 
                   + {1\over 32}\,{1\over \yp}
                  \bigg)
              \bigg]
\nonumber\\&&\mbox{} 
           + \bigg[
              {75857\over 466560} 
              + {7\over 32}\,{1\over \yp} 
              - {121895171\over 204120000}\,\yp 
              + {4262581\over 10206000}\,\yp^2 
              - {3177101006\over 4011170625}\,\yp^3
\nonumber\\&&\mbox{}
\hspace{1em}
             + \zeta_2\,\bigg(
                  {173\over 1296} 
                  + {67\over 2592}\,{1\over \yp}
                  + {53\over 324}\,\yp 
                  \bigg) 
              + \zeta_3\,\bigg(
               - {1\over 18}\,{1\over \yp}
               + {7\over 90}\,\yp 
               - {1\over 45}\,\yp^{2} 
               - {16\over 315}\,\yp^3 
              \bigg)
\nonumber\\&&\mbox{}
\hspace{1em}
              + \logtwos\,\bigg(
                  - {757\over 7776} 
                  - {331\over 7776}\,{1\over \yp}
                  - {95\over 3888}\,\yp 
                  \bigg) 
              + \logctheta^2\,\bigg(
                  - {103\over 2592} 
                  - {1\over 81}\,{1\over \yp}
                  + {1\over 300}\,\yp 
\nonumber\\&&\mbox{}
\hspace{1em}
                  - {103\over 1080}\,\yp^2 
                  - {5314\over 33075}\,\yp^3 
                  \bigg) 
              + \logctheta\,\bigg(
                  - {527\over 7776} 
                  + {11\over 288}\,{1\over \yp}
                  - {1489\over 30375}\,\yp 
                  + {1081\over 9720}\,\yp^2 
\nonumber\\&&\mbox{}
\hspace{1em}
                  - {3338578\over 10418625}\,\yp^3 
                  \bigg)
              + {\cal O} (\yp^4)
              \bigg] 
\nonumber\\&&\mbox{}
           + {M_W^2\over M_t^2}\,\bigg[
               - {384287\over 1555200} 
               + {47\over 3456}\,{1\over \yp} 
               - {5741993\over 5443200}\,\yp 
               - {217997\over 226800}\,\yp^2 
               + {149\over 1215}\,\yp^3 
               + {5519\over 110565}\,\yp^4 
\nonumber\\&&\mbox{}
\hspace{1em}
               + \logtwos\,\bigg(
                   - {83083\over 155520} 
                   - {1819\over 5184}\,{1\over \yp}
                   + {15017\over 38880}\,\yp 
                   + {3977\over 38880}\,\yp^2 
                  \bigg) 
               + \logctheta\,\bigg(
                   - {3343\over 155520} 
                   - {7\over 1296}\,{1\over \yp}
\nonumber\\&&\mbox{}
\hspace{1em}
                   - {823\over 38880}\,\yp 
                   + {17\over 38880}\,\yp^2 
                  \bigg)
               + \zeta_2\,\bigg(
                     {257\over 864} 
                   + {175\over 864}\,{1\over \yp}
                   + {13\over 144}\,\yp 
                   + {11\over 18}\,\yp^2 
                  \bigg)
               + {\cal O} (\yp^5)
              \bigg] 
\nonumber\\&&\mbox{}
           + \left({M_W^2\over M_t^2}\right)^{2}\,\bigg[
                 {470213\over 1166400} 
               - {6119\over 31104}\,{1\over \yp} 
               + {54467207\over 105840000}\,\yp 
               - {640160203\over 158760000}\,\yp^2 
               - {3791409547\over 982327500}\,\yp^3 
\nonumber\\&&\mbox{}
\hspace{1em}
               - {44597\over 173745}\,\yp^4 
               - {38284\over 405405}\,\yp^5 
               + \zeta_2\,\bigg(
                   - {239\over 324} 
                   + {10\over 27}\,{1\over \yp}
                   - {14539\over 10800}\,\yp 
                   + {4769\over 1080}\,\yp^2 
\nonumber\\&&\mbox{}
\hspace{1em}
                   + {4838\over 2025}\,\yp^3 
                  \bigg)
               + \logtwos\,\bigg(
                   - {14791\over 38880} 
                   - {635\over 864}\,{1\over \yp}
                   + {2935841\over 1701000}\,\yp 
                   - {1769611\over 1701000}\,\yp^2 
                   - {19039\over 70875}\,\yp^3
                   \bigg) 
\nonumber\\&&\mbox{}
\hspace{1em}
                + \logtwos\,\logctheta\,\bigg(
                   {1\over 36} 
                   + {1\over 144}\,{1\over \yp}
                   + {1\over 36}\,\yp 
                   \bigg) 
               + \logctheta\,\bigg(
                   - {623\over 19440} 
                   - {7\over 864}\,{1\over \yp}
                   - {105443\over 3402000}\,\yp 
\nonumber\\&&\mbox{}
\hspace{1em}
                   + {283\over 212625}\,\yp^2 
                   - {31\over 283500}\,\yp^3 
                  \bigg)
               + \logtwos^2\,\bigg(
                     {1\over 36} 
                   + {1\over 144}\,{1\over \yp}
                   + {1\over 36}\,\yp 
                  \bigg)
               + {\cal O} (\yp^6)
              \bigg] 
\nonumber\\&&\mbox{}
           + \left({M_W^2\over M_t^2}\right)^{3}\,\bigg[
                 {1103711\over 388800} 
               - {165325\over 373248}\,{1\over \yp} 
               + {608169487\over 190512000}\,\yp 
               - {91635792023\over 6429780000}\,\yp^2 
\nonumber\\&&\mbox{}
\hspace{1em}
               - {345910967113\over 17681895000}\,\yp^3 
               - {3664430299069\over 229864635000}\,\yp^4 
               + {655346\over 1216215}\,\yp^5 
               + {238846\over 1342575}\,\yp^6
\nonumber\\&&\mbox{}
\hspace{1em}
               + \zeta_2\,\bigg(
                   - {9497\over 2592} 
                   + {1397\over 2592}\,{1\over \yp}
                   - {6163\over 3240}\,\yp 
                   + {52021\over 3240}\,\yp^2 
                   + {163\over 25}\,\yp^3 
                   + {6392\over 675}\,\yp^4 
                  \bigg) 
\nonumber\\&&\mbox{}
\hspace{1em}
               + \logtwos\,\bigg(
                     {233569\over 155520} 
                   - {36293\over 31104}\,{1\over \yp}
                   + {1328861\over 453600}\,\yp 
                   - {169446869\over 20412000}\,\yp^2 
                   + {1726367\over 637875}\,\yp^3 
\nonumber\\&&\mbox{}
\hspace{1em}
                   + {62576\over 91125}\,\yp^4 
                   \bigg)                   
               + \logtwos\, \logctheta\,\bigg(
                   {5\over 81} 
                   + {5\over 324}\,{1\over \yp}
                   + {5\over 81}\,\yp 
                   \bigg) 
               + \logctheta\,\bigg(
                   - {5419\over 155520} 
                   - {11\over 1296}\,{1\over \yp}
\nonumber\\&&\mbox{}
\hspace{1em}
                   - {51157\over 1360800}\,\yp 
                   - {78299\over 20412000}\,\yp^2 
                   - {148\over 637875}\,\yp^3 
                   + {7\over 182250}\,\yp^4 
                  \bigg)
\nonumber\\&&\mbox{}
\hspace{1em}
               + \logtwos^2\,\bigg(
                     {5\over 81} 
                   + {5\over 324}\,{1\over \yp}
                   + {5\over 81}\,\yp 
                  \bigg)
               + {\cal O} (\yp^7) 
              \bigg]
\nonumber\\&&\mbox{}
          + {\cal O} \left( \left( {M_W^2\over M_t^2} \right)^4 \right)
\bigg\}\,.
\label{eqzbbOS2}
\end{eqnarray}
For the definition of the top quark mass, $M_t$, the
on-shell scheme has been adopted. With the help of the equation
\begin{eqnarray}
  M_t &=& m_t(\mu) \left[1+ \frac{\alpha_s(\mu)}{\pi} C_F
  \left(1+\frac{3}{4}\ln\frac{\mu^2}{m_t^2(\mu)}\right) \right]
\label{eqostoms}
\end{eqnarray}
the transformation to the $\overline{\rm MS}$ scheme can be performed.

There are several checks for the correctness of our results.
First of all, our new calculation with the automated HMP provides
an independent check of the $M_t^2$ and the $\ln M_t^2$ term.
The terms up to (and including) order $1/M_t^2$
and $1/c_\Theta^2$ were calculated with arbitrary gauge parameters 
$\xi_W$ and $\xi_S$ for the electroweak sector and QCD, respectively. 
We could check that to each order in the 
$1/M_t$ expansion $\xi_W$ drops out separately. Whereas for the 
$M_t^2$ and $1/M_t^2$ terms this happens after taking the sum of 
all contributing diagrams, for the $(M_t^2)^0$ order only 
the difference 
$\delta\Gamma_b^{0,W} - \delta\Gamma_d^{0,W}$
renders the result gauge invariant\footnote{
  It should be noted that we partly had to repeat the calculation done
  in $\cite{CzaKue96}$ for arbitrary gauge parameter $\xi_W$.}.
$\xi_S$ already drops out if only the sum of diagrams contributing
to the separate classes (see Fig.~\ref{fig2l}) is considered.
The $1/M_t^4$ and $1/M_t^6$ corrections and the higher order terms
in the $1/c_\Theta^2$ expansion were computed in Feynman gauge
as then the calculation is much faster. 
Note, that in Eq.~(\ref{eqzbbOS2}) where all parameters are expressed 
in the on-shell scheme the explicit $\mu$ dependence drops out
which is also a welcome check for our calculation.

Let us now discuss the numerical relevance of the newly computed terms.
Using $s_\Theta^2=0.223$ the corrections induced by $W$-bosons read in
the on-shell scheme:
\begin{eqnarray}
\lefteqn{\delta\Gamma^W_{b-d} = \Gamma^0
  {1\over s_\Theta^2}{\alpha\over \pi}\times}\nonumber\\&&
\bigg\{
       - 0.11\,{M_t^2\over M_W^2}
       + 0.71 
       - 0.31\,\logtwos 
       + \left(
             0.36
           - 0.89\,\logtwos
          \right)\,{M_W^2\over M_t^2} 
\nonumber\\&&\mbox{}
       + \left(
           - 0.24
           - 0.97\,\logtwos
          \right)\,\left({M_W^2\over M_t^2}\right)^{2} 
       + \left(
           - 0.78
           - 0.43\,\logtwos
          \right)\,\left({M_W^2\over M_t^2}\right)^{3} 
\nonumber\\&&\mbox{}
       + {\alpha_s\over \pi}\,\bigg[
             0.24\,{M_t^2\over M_W^2}
           + 1.21
           - 0.32\,\logtwos 
           + \left(
                 1.40
               - 1.99\,\logtwos
              \right)\,{M_W^2\over M_t^2} 
\nonumber\\&&\mbox{}\hspace{1em}
           + \left(
                 0.37
               - 2.99\,\logtwos 
               + 0.08\,\logtwos^2
              \right)\,\left({M_W^2\over M_t^2}\right)^{2} 
\nonumber\\&&\mbox{}\hspace{1em}
           + \left(
               - 1.08
               - 2.64\,\logtwos 
               + 0.17\,\logtwos^2
              \right)\,\left({M_W^2\over M_t^2}\right)^{3} 
          \bigg] 
\bigg\}
+ {\cal O} \left( \left( {M_W^2\over M_t^2} \right)^4 \right)\,.
\label{eqdelWnum}
\end{eqnarray}
Using Eq.~(\ref{eqostoms}) one may transform the top mass, $M_t$, to the
$\overline{\rm MS}$ scheme. For the renormalization scale $\mu$
explicitly appearing in the resulting expression,
$\delta\bar\Gamma^W_b$, we adopt the choice $\mu^2=M_Z^2$.
One finds ($m_t$ denotes the $\overline{\rm
  MS}$ top mass, and $l_{tW} = \ln(m_t^2/M_W^2)$):
\begin{eqnarray}
\lefteqn{\delta\bar\Gamma^W_{b-d} = \Gamma^0
  {1\over s_\Theta^2}{\alpha\over \pi}\times}\nonumber\\&&
\bigg\{
       - 0.11\,{m_t^2\over M_W^2}
       + 0.71 
       - 0.31\,\logtwms 
       + \left(
             0.36
           - 0.89\,\logtwms
          \right)\,{M_W^2\over m_t^2} 
\nonumber\\&&\mbox{}
       + \left(
           - 0.24
           - 0.97\,\logtwms
          \right)\,\left({M_W^2\over m_t^2}\right)^{2} 
       + \left(
           - 0.78
           - 0.43\,\logtwms
          \right)\,\left({M_W^2\over m_t^2}\right)^{3} 
\nonumber\\&&\mbox{}
       + {\alpha_s\over \pi}\,\bigg[
             \left(
               - 0.09
               + 0.21\,\logtwms
              \right)\,{m_t^2\over M_W^2}
           + 0.24
           + 0.30\,\logtwms 
\nonumber\\&&\mbox{}\hspace{1em}
           + \left(
               - 2.57
               + 3.34\,\logtwms 
               - 1.78\,\logtwms^2
              \right)\,{M_W^2\over m_t^2} 
\nonumber\\&&\mbox{}\hspace{1em}
           + \left(
               - 1.16
               + 4.12\,\logtwms 
               - 3.80\,\logtwms^2
              \right)\,\left({M_W^2\over m_t^2}\right)^{2} 
\nonumber\\&&\mbox{}\hspace{1em}
           + \left(
                 4.99
               - 2.37\,\logtwms 
               - 2.41\,\logtwms^2
              \right)\,\left({M_W^2\over m_t^2}\right)^{3} 
          \bigg]
\bigg\}
+ {\cal O} \left( \left( {M_W^2\over m_t^2} \right)^4 \right)\,.
\end{eqnarray}
One observes that for realistic values of $M_Z$ and $M_t$, respectively
$m_t$, the constant at next-to-leading order dominates over the 
$\ln M_t^2$ term known before.
Furthermore, the size of the contributions from the individual
$1/M_t^2$ orders is roughly the same in both schemes.  It is remarkable
that at one-loop level the corrections arising from the $1/M_t^2$ terms are 
of similar size than the one from next-to-leading order, however,
the signs are different. The higher order corrections in $1/M_t$
are smaller, which means that effectively only the leading $M_t^2$ term
remains.
Proceeding to two loops the situation is similar: Starting at
${\cal O}(1/M_t^2)$ the sign is opposite as compared to the leading
terms and a large cancellation takes place. Here, the $1/M_t^4$
term is still comparable with the $1/M_t^2$ contribution.
The $1/M_t^6$ term, however, is small and thus
strongly suggests that the presented terms should provide a very good
approximation to the full result.

For $M_t=175$~GeV, $M_Z=91.19$~GeV, $\alpha = 1/129$ and
$\alpha_s(M_Z)=0.120$ we get:
\begin{eqnarray}
  \delta\Gamma_{b-d}^W &=& (-5.69 - 0.79 + 0.50 + 0.06)\mbox{ MeV}
\,\,=\,\, -5.92\mbox{ MeV} \,.
\label{eqdelwnum}
\end{eqnarray}
The first two numbers in Eq.~(\ref{eqdelwnum}) correspond to the 
${\cal O}(\alpha)$, the second two to the ${\cal O}(\alpha\alpha_s)$
corrections.
Each of these contributions is again separated into the $M_t^2$ terms and the
sum of the subleading ones.  For the diagrams containing a $Z$ boson
\cite{CzaKue96} the numerical value reads:
\begin{eqnarray}
  \delta\Gamma_b^Z &=& (0.52 - 0.01)\mbox{ MeV} \,\,=\,\,
  0.51\mbox{ MeV}\,.
\label{eqZboson}
\end{eqnarray}
One can see that even the contributions of the subleading terms 
of diagrams arising from a $W$ boson exchange are more important
than the corrections resulting from $\delta\Gamma_b^Z$.

Let us finally also give the net effect of the non-factorizable
corrections. Adopting a notation analogue to Eq.~(\ref{eqgamdiff}) and
using the expansion for $\delta\Gamma_d^W$ in the limit of small $\yp$
\cite{CzaKue96}\footnote{We had to increase the depth of the expansion
  in $\yp$ for these terms.}
the difference to the
result where ``naive'' factorization is assumed reads
\begin{eqnarray}
{\alpha_s\over \pi} \left(
C_F\,\delta\Gamma_{b}^{(2),W} - \delta\Gamma_{b}^{(1),W}\right) &=&
0.68 \mbox{ MeV}\,.
\label{eqfac1}
\end{eqnarray}
This is comparable both to the error on the present experimental value
of $\Gamma_{\rm had}$ and even to the one-loop corrections from the $Z$
boson diagrams (see Eq.~(\ref{eqZboson})).  For $\Gamma(Z\to b\bar{b})$
the factorization is often performed such that the leading $M_t^2$ term
is reproduced correctly.  If this is taken into account the deviation
corresponding to Eq.~(\ref{eqfac1}) reduces to $-0.04$~MeV which is
still larger than the mixed ${\cal O}(\alpha\alpha_s)$ corrections
induced by internal $Z$ bosons (second number in Eq.~(\ref{eqZboson})).

To conclude, the missing non-universal piece to the  decay of the
$Z$ boson into bottom quarks to ${\cal O}(\alpha\alpha_s)$
has been computed. 
An expansion for large top quark mass has been
performed and it was demonstrated that only the inclusion of
power suppressed terms in $1/M_t^2$ leads to reliable
predictions. The results are presented in a form which allows
a simple implementation into program libraries for the description of
the $Z$ line shape (see \cite{YelRep95} and references therein).

\vspace{5mm} 
\centerline{\bf Acknowledgments} 
\medskip
\noindent 
We want to thank K.G.~Chetyrkin and J.H.~K\"uhn for fruitful discussions
and careful reading of the manuscript.  We are grateful to A.~Czarnecki
for helpful discussions and for providing us with the results of some
individual diagrams from \cite{CzaKue96} for comparison.  The
discussions with P.~Gambino, W.~Hollik and G.~Weiglein are greatly
acknowledged. The work of R.H. was supported by the
``Landesgraduiertenf\"orderung'' and the ``Graduiertenkolleg
Elementarteilchenphysik'' at the University of Karlsruhe. This work was
supported by DFG Contract Ku 502/8-1.


\end{document}